\documentclass{article}
\usepackage[utf8]{inputenc}
\pdfoutput=1
\usepackage{jcapmod} 
\usepackage{booktabs} 
\usepackage[english]{babel}												
\usepackage{amsmath,amssymb,amsbsy,amstext, amsthm, simplewick}
\usepackage{wrapfig}
\usepackage{hyperref}
\usepackage{graphicx}
\usepackage{amsfonts}
\usepackage{amssymb}
\usepackage{subfig}
\usepackage{enumitem}
\usepackage{upgreek}
\usepackage{exscale,relsize}
\usepackage[makeroom]{cancel}
\usepackage{soul}
\usepackage{bbold}
\usepackage{lipsum}
\usepackage{mdframed}
\usepackage{mathtools}
\usepackage[export]{adjustbox}
\usepackage[outdir=./]{epstopdf}
\usepackage{placeins}
\usepackage{afterpage}
\usepackage{float}
\usepackage{comment}

\newcommand{\ud}{\mathrm{d}}

\newcommand{\Beq}{\begin{equation}\begin{aligned}}
\newcommand{\Eeq}{\end{aligned}\end{equation}}
\newcommand{\g}{g_{a\gamma}}
\newcommand{\GeV}{{\rm GeV}}

\begin{document}

\hypersetup{pageanchor=false}
\begin{titlepage}
	
\setcounter{page}{1} \baselineskip=15.5pt \thispagestyle{empty}
	
\bigskip\
	
\vspace{1cm}
\begin{center}
		
{\fontsize{20.74}{24}\selectfont  \sffamily \bfseries  Electromagnetic Bursts from Mergers of Oscillons in Axion-like Fields}

\end{center}
	
\vspace{0.2cm}
	
\begin{center}
{\fontsize{12}{30} Mustafa A. Amin\footnote{mustafa.a.amin@rice.edu}, Zong-Gang Mou\footnote{zm17@rice.edu}}
\end{center}
\begin{center}
		
\vskip 7pt
		
\textsl{Department of Physics \& Astronomy, Rice University, Houston, Texas 77005, U.S.A.}\\

\vskip 7pt
		
\end{center}
	
\vspace{1.2cm}
\hrule \vspace{0.3cm}
\noindent {\sffamily \bfseries Abstract} \\[0.1cm]
We investigate the bursts of electromagnetic and scalar radiation resulting from the collision, and merger of oscillons made from axion-like particles using 3+1 dimensional lattice simulations of the coupled axion-gauge field system.   The radiation into photons is suppressed before the merger. However, it becomes the dominant source of energy loss after the merger if a resonance condition is satisfied. Conversely, the radiation in scalar waves is dominant during initial merger phase but suppressed after the merger. The backreaction of scalar and electromagnetic radiation is included in our simulations. We evolve the system long enough to see that the resonant photon production extracts a significant fraction of the initial axion energy, and again falls out of the resonance condition. We provide a parametric understanding of the time, and energy scales involved in the process and discuss observational prospects of detecting the electromagnetic signal. 

\vskip 10pt
\hrule
\vskip 10pt

\vspace{0.6cm}
\end{titlepage}

\hypersetup{pageanchor=true}

\tableofcontents

\newpage
\section{Introduction}
Axions and axion-like particles are well motivated candidates for dark matter \cite{Wilczek:1977pj,Peccei:1977hh,Preskill:1982cy,Abbott:1982af,Dine:1982ah,Ringwald:2014vqa,Marsh:2015xka}. Like other dark matter candidates, their non-gravitational interactions with the Standard Model (SM) have never been detected \cite{Graham:2015ouw,Irastorza:2018dyq}. Axions naturally couple to photons via the $\g\phi F_{\mu\nu}\tilde{F}^{\mu\nu} $ interaction where $\phi$ is the axion field, $F_{\mu\nu}$ is the electromagnetic field strength tensor, and $\tilde{F}_{\mu\nu}$ is its dual. Strong constraints exist on $\g$ from astrophysical observations and terrestrial experiments \cite{Tanabashi:2018ocaf}. However, under certain circumstances, even a very feeble interaction with photons can give rise to dramatic effects. Explosive production of photons due to parametric resonance from coherently oscillating axion fields is one such scenario.

Gravitational clustering, and attractive self-interactions can cause axion-like fields to form coherently oscillating, spatially localized, metastable, solitonic configurations (for early work, see \cite{Hogan:1988mp,Kolb:1993hw}, and also see, for example, \cite{Amin:2011hj,Schive:2014dra,Amin:2019ums}). If gravitational interactions are ignored, such configurations are called oscillons, and have a long history (see, for example, \cite{Bogolyubsky:1976yu,Gleiser:1993pt,Copeland:1995fq,Kasuya:2002zs,Amin:2010jq,Zhang:2020bec}). If gravity and self-interactions are relevant, and we focus on cosine potentials, these configurations are called  ``axion stars", which can be dilute or dense (see, for example, \cite{Eby:2019ntd,Visinelli:2017ooc}). In dilute axion stars, self-interaction can typically be ignored compared to gravity, whereas in dense stars, self-interactions play a dominant role in the dynamics of the axion star. 
In this work, we will focus on oscillons in axion-like fields as a source for resonantly producing photons. We will consider a broader class of potentials that flatten away from the minimum instead of considering the periodic cosine potential. 

\begin{figure}[t]
	\centering
	\includegraphics[width=1.01\linewidth]{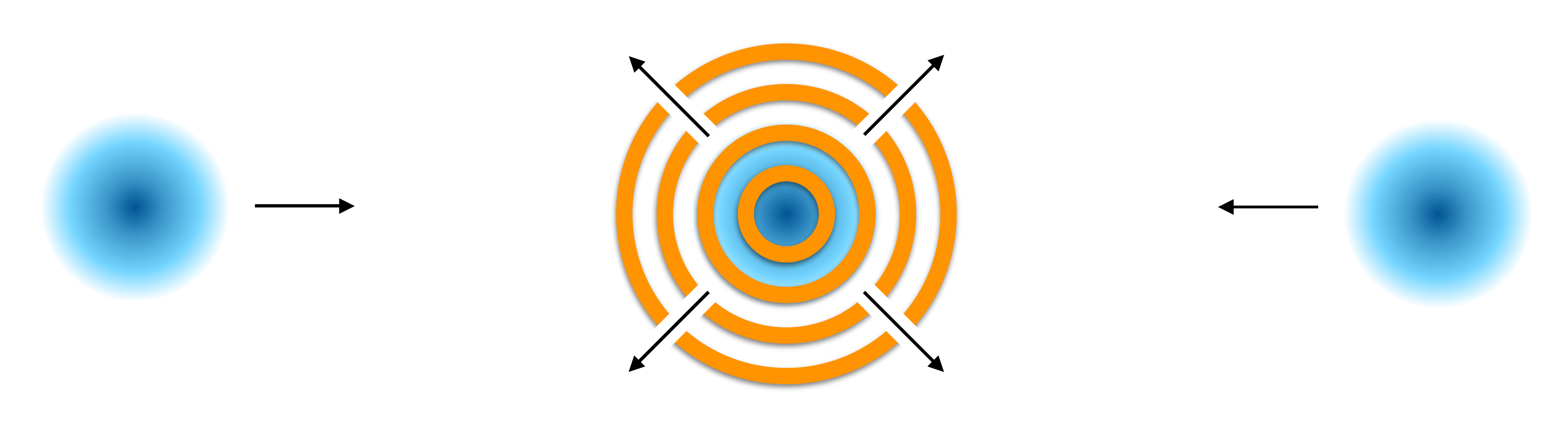}
	\caption{A schematic picture of resonant electromagnetic field production triggered by the merger of spatially localized, coherent field configurations in axion-like fields (oscillons).}
	\label{fig:CartoonEMBurst}
\end{figure}

Resonant phenomena converting axions to photons depend on the amplitude and coherence length of the oscillating axion field. Since, the field amplitude in the centre of oscillons can be large, and does not redshift with the expansion of the universe, oscillons can provide a natural source for resonant particle production under certain conditions. The presence of such configurations in the early  or present day universe creates a possibility of particle production from localized regions of space -- quite distinct from resonance from a homogeneous oscillating field. However, there has to be a trigger to start off this production because if the condition for resonance is always satisfied, then these oscillons would have already decayed away. A possible trigger is a collision and merger of oscillons. We study such head-on collisions, mergers, and associated resonant photon production in this paper using full $3+1$ dimensional lattice simulations of the axion-photon system (see Fig.~\ref{fig:CartoonEMBurst}).

The recent results in \cite{Hertzberg:2018zte,Levkov:2020txo} motivated us to pursue our present work, along with related earlier work \cite{Kephart:1995,Tkachev:2014dpa}.\footnote{MA also acknowledges a short collaboration with D. Grin and M. Hertzberg almost 11 years ago (unpublished work) regarding resonant photon axion conversion in localized axion clumps.} In these works, a QCD axion was the prime focus, and they were mainly concerned with the dilute branch of axion stars or more broadly, just a localized gravitationally bound clump of axions. In this regime, the authors in \cite{Hertzberg:2018zte} in particular, provides a condition for resonance to be effective for a single axion star. In \cite{Hertzberg:2020dbk}, the authors then numerically investigated the collision of non-relativistic (dilute) axion stars to understand the merger process, albeit without coupling to photons for the simulations. They showed that two axion stars which do not satisfy the resonance condition before merger, can do so after the merger. The authors in \cite{Levkov:2020txo} also carried out elegant analytical and numerical work for resonant photon production, but without full collision dynamics or including backreaction in the simulations.

In the current work, we focus on a more general class of axion potentials, specifically, non-periodic potentials that flatten as we move away from the minimum. Such potentials are well-motivated theoretically \cite{Silverstein:2008sg, McAllister:2014mpa, Kallosh:2013hoa}, and have been pursued as modeling inflation and its end \cite{Amin:2011hj, Lozanov:2017hjm}, dark matter \cite{Amin:2011hu,Olle:2019kbo,Soda:2017dsu} and dark energy \cite{Amin:2011hu,Garcia-Garcia:2018hlc}. Unlike the usual cosine potential, or quadratic potentials with or without quartic interactions, our flattened potentials support very long-lived oscillon configurations even without gravitational interactions. Depending on the  shape of the potential, isolated oscillons can last for $\gtrsim 10^{9}$ oscillations (see, for example, \cite{Zhang:2020bec}).

 In contrast to earlier work, we use $3+1$ dimensional lattice simulations to solve the axion-photon system (using link-variables\cite{smit2002introduction}). We do not resort to any non-relativistic\footnote{In the special-relativistic sense.} approximations for the axion field either, and include the effects of strong self-interactions. For some parameter regimes, we are able to simulate the entire process from pre-collision, to the shutting down of resonance due to backreaction of the gauge fields. Starting with pre-collision oscillons that do not yield resonant photon production, we track the axion and photon fields through the collision, and finally through the phase of resonant gauge field production which eventually self-regulates the oscillon and makes it fall out of the resonance condition. We keep tract of both scalar and gauge field radiation throughout the process. 

Our work has some important limitations, partly because we have not included gravitational interactions in the simulations -- they are discussed in Section ~\ref{sec:Limitations}. Recognizing these limitations, we proceed to investigate the full nonlinear dynamics of the merger of oscillons, photon production and backreaction. We will provide a parametric understanding of the process, and hope that the lessons learnt here can be applied much more broadly. In future work we plan to include gravitational interactions, both in the dilute axion-star regime, as well as the strong-field gravity limit in our simulations. We also note that while we continue referring to the gauge fields as the electromagnetic field/photon-field, our work can also applied to any gauge fields (for example, dark photons \cite{Agrawal:2018vin})  and can also be applied to study similar phenomenon in the early universe, for example at the end of inflation or during some other phase transition \cite{Amin:2011hj,Amin:2019ums,Amin:2018kkg,Musoke:2019ima,Adshead:2019lbr,Fukunaga:2020mvq}.

The rest of the paper is organized as follows. In Sec.~\ref{sec:Theoretical}, we provide the continuum as well as discretized versions of action for the axion-photon system. In Sec.~\ref{sec:Analytical}, we provide a brief overview of oscillons in these theories, as well as the conditions and time-scales related to the resonant production of photons from oscillons. In Sec.~\ref{sec:Collisions} we discuss the collision and merger of oscillons. In Sec.~\ref{sec:EMBurst} we focus on the resonant production of photons from the post-merger oscillon, as well as the backreaction of this photon production on oscillons. We briefly review possible phenomenological/observational implications of resonant photon production in Sec.~\ref{sec:Phenomenology}. We discuss the main limitations of our work, and some related future directions in Sec.~\ref{sec:Limitations}. We end in Sec.~\ref{sec:Summary} with a summary of our results. In two appendices, we provide further details of our numerical simulations, including details of initial conditions, discretization and evolution schemes, and a wider exploration of the parameter space.

\section{The Theoretical Setup and Numerical Algorithm}
\label{sec:Theoretical}
In this section we provide the underlying action and equations of motion for our system of interest, as well as the basics of the discretization and time evolution algorithm to numerically evolve the system. The reader who is not interested in numerical details can skip Section \ref{sec:Numerical} entirely.
More details of our explicit model for the axion field potential, and analytic considerations are provided in Section \ref{sec:Analytical}. Some more technical aspects of the numerical algorithm, and the setting up of initial conditions on the lattice, are relegated to the appendix. 

\subsection{Action and Equations of Motion}
Our system consists of a pseudo-scalar field $\phi$ coupled to the electromagnetic field. The action for our system is given by
\Beq
\label{eq:action_c}
S=\int d^4 x\left[-\frac{1}{2}\partial_\mu\phi\partial^\mu\phi-V(\phi)-\frac{1}{4}F_{\mu\nu}F^{\mu\nu}-\frac{\g}{4}\phi F_{\mu\nu}\tilde{F}^{\mu\nu}\right]\,,
\Eeq
where we adopt $-+++$ signature of the metric. The electromagnetic field-strength tensor is 
$F_{\mu\nu}=\partial_\mu A_\nu-\partial_\nu A_\mu$, and $\tilde{F}^{\mu\nu}$ is the dual field strength tensor. The equations of motion for the axion and the gauge fields are given by 
\Beq
&\partial_\mu\partial^\mu\phi-\partial_\phi V=\frac{\g}{4} F_{\mu\nu}\tilde{F}^{\mu\nu}\,,\\
&\partial_\mu F^{\mu\nu}=-\g\partial_\mu\phi \tilde{F}^{\mu\nu}\,.
\Eeq
In terms of the electric and magnetic fields $E_i=F_{i0}$ and $B_i=(1/2)\epsilon_{ijk}F_{jk}$, we can write the above equations as 
\Beq
&\partial_\mu\partial^\mu\phi-\partial_\phi V=-\g {\bf E}\cdot{\bf B}\,
,
\\
&\partial_0\left( {\bf E} +\g \phi  {\bf B} \right) - {\bf \nabla} \times \left( {\bf B} -\g \phi  {\bf E} \right)=0
,\quad
{\bf \nabla} \cdot \left( {\bf E} +\g \phi  {\bf B} \right)=0\,
.
\Eeq

\subsection{Numerical Algorithm}
\label{sec:Numerical}
We discretize the action \eqref{eq:action_c} on a space-time lattice as follows \cite{smit2002introduction}:
\begin{align}
S=& \sum_{x}\Bigg[
\frac{1}{2}\left(\frac{\phi(x+\ud t)-\phi(x)}{\ud t}\right)^2
-\sum_i\frac{1}{2}\left(\frac{\phi(x+\ud x_i)-\phi(x)}{\ud x_i}\right)^2
-V\left(\phi\right) +
\nonumber \\ & 
\sum_i \frac{2}{(\ud t\ud x_i)^2} \left(2-U_{0i} - U_{i0}\right)
 -\sum_{ij}\frac{1}{(\ud x_i \ud x_j)^2} \left(2-U_{ij} - U_{ji}\right)
\Bigg]
+S_1,
 \end{align}
where $S_1$ is the interaction part of the action that we will specify shortly. The $U_{\mu\nu}$ is the lattice ``plaquette", defined as a product of ``gauge links" $U_\mu$:
\Beq
U_{\mu\nu}(x)= U_\mu(x) U_\nu(x+\ud x_\mu) U^\dagger_\mu(x+\ud x_\nu) U^\dagger_\nu(x)\,.
\Eeq
By assuming $U_\mu(x) = \exp \left(\frac{i}{2}\ud x_\mu A_\mu(x)\right)$, we can recover the continuum action in the limit $\ud x_\mu \to 0$ (no Einstein summation). Note that $U_{\mu\nu}=\exp(\frac{i}{2}\ud x_\mu \ud x_\nu F_{\mu\nu}+{\rm corrections})$.

The challenging aspect is to appropriately discretize the interaction part of the action:
\begin{align}
S_1= 8\pi^2g_{a\gamma} \int \ud^4 x   \phi \left(-\frac{1}{64\pi^2}\right)\epsilon^{\mu\nu\rho\sigma}F_{\mu\nu}F_{\rho\sigma}
.
\end{align}
Note that (without the $\phi$), the above term is the Chern-Simons number. It is known that a naive use of the plaquettes here yields results close to the continuum expectation at leading order, but can produce large additional contributions at the next order in the $dx_\mu$s. An improved discretization of this term (as
implemented by \cite{Tranberg:2003gi,DiazGil:2008tf,Mou:2017zwe,Mou:2018xto}) is as follows:
\begin{align}
\left(-\frac{1}{64\pi^2}\right)\epsilon^{\mu\nu\rho\sigma}F_{\mu\nu}F_{\rho\sigma}
= \left(-\frac{1}{2\pi^2 \ud^4x}\right)\left(I_{01}I_{23}+ I_{02}I_{31}+ I_{03}I_{12} \right)
,
\end{align}
with
\begin{align}
I_{\mu \nu}(x) = \frac{1}{4}{\rm Im} \left[
U_{\mu\nu}(x)
+U_{\mu\nu}(x-\ud x_\mu)
+U_{\mu\nu}(x-\ud x_\mu-\ud x_\nu)
+U_{\mu\nu}(x-\ud x_\nu)
\right].
\end{align}
This choice of discretization leads to significant suppression of corrections beyond leading order in $\ud x_\mu$s. We refer the reader to \cite{Tranberg:2003gi,Mou:2017zwe} for more details. 

The discretized versions of the  equations of motion for $\phi$ and $U_i$ can then be derived from the derivation of the action with respect to the fields in a straightforward manner. These equations are provided explicitly in Appendix \ref{sec:AppNumerical}. We note that the presence of Chern-Simons term makes it difficult to solve the equations of motion on the lattice because the evolution equations become implicit. These equations follow a general pattern
\begin{align}
\label{eq:eq}
X = \beta+ \alpha F(X),
\end{align}
where $X$ stands for $\dot\phi$ and $E_i$, and $\beta$ and $\alpha$ are constant. Here, $F$ depends on $X$ in a complicated way. Instead of solving these equations algebraically, it is easier to solve the recurrence equation:
\begin{align}
\label{eq:re}
X_{n+1} = \beta+ \alpha F(X_n),
\end{align}
of which $X$ is a fixed point. So the question of solving $X$ in (\ref{eq:eq}) becomes that of how to reach the fixed point $X$ in (\ref{eq:re}). As long as the parameter $\alpha$, which is proportional to $\ud t$ and $g_{a\gamma}$, is tiny, the fixed point is an attractive one, and we can reach $X$, starting from $\beta$, within a few iterations. More details on the lattice setup, including initial conditions can be found in Appendix \ref{sec:AppNumerical}.

Use of gauge links automatically leads to the preservation of gauge invariance in the discrete theory, and constraint equations (Gauss's law) are naturally respected by the evolution. These features are lost in other discretization and (explicit) evolution schemes, however they can still be used as long as constraint equations are satisfied to the required precision. For an explicit (as opposed to implicit) approach to solving the axion-gauge field system, see for example \cite{Adshead:2015pva,Figueroa:2017qmv,Figueroa:2020rrl}. For an explicit,  symplectic, link-variable based approach for a charged scalar+gauge field system in a self-consistently expanding universe, see \cite{Lozanov:2019jff}.

\section{Analytic Considerations}
\label{sec:Analytical}
\subsection{Model for the Axion-like Field}
\begin{figure}[t]
	\centering
	\includegraphics[width=1.01\linewidth]{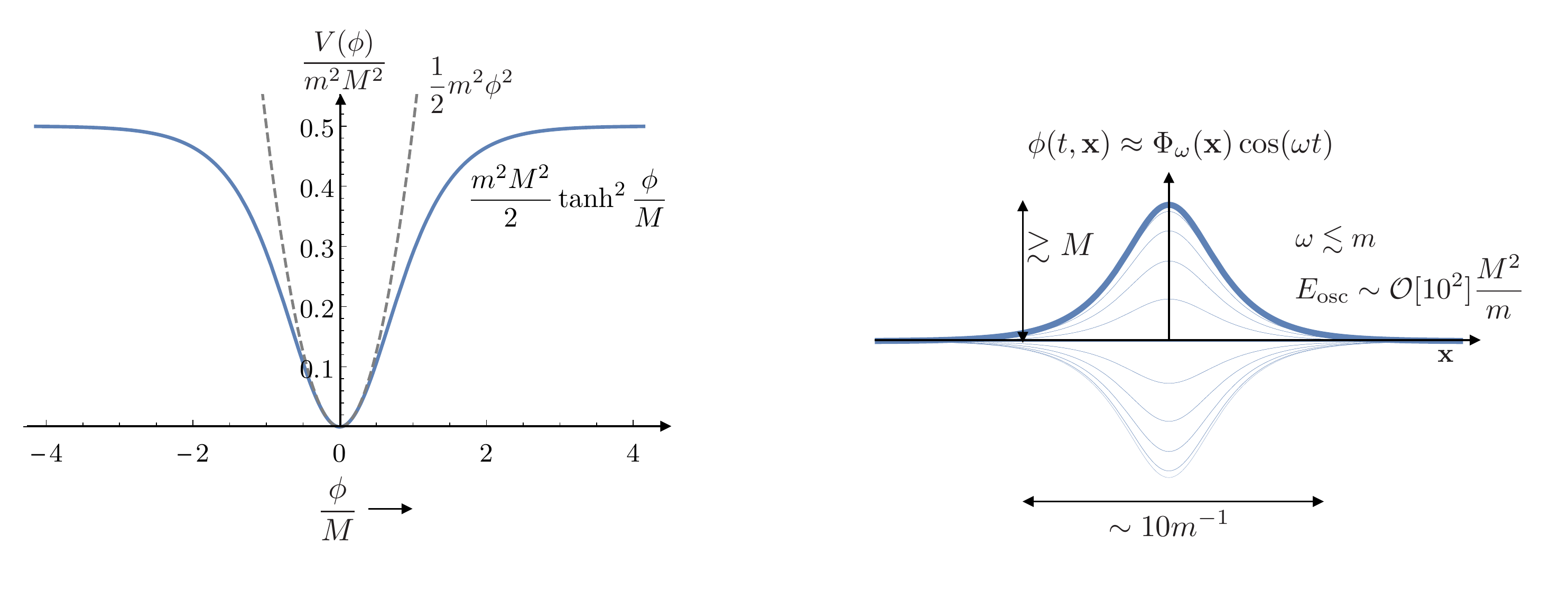}
	\caption{A flattened scalar field potential (left) which supports long-lived, spatially localized field configurations called oscillons (right).}
	\label{fig:Potential}
\end{figure}

For the scalar field potential, $V(\phi)$, we will assume that
\Beq
V(\phi)=m^2M^2\,U(\phi/M)\quad \textrm{where}\quad U(x\ll 1)=(1/2)x^2+\hdots\quad\textrm{and}\quad U(x\gg 1)\propto x^{\alpha<2}.
\Eeq
Such potentials provide sufficient attractive self-interaction to support oscillons (discussed below). As a concrete example, we will consider a potential of the form
\Beq
V(\phi)=\frac{m^2M^2}{2}\tanh^2\left(\frac{\phi}{M}\right)\,.
\Eeq
Note that $M$ plays the role of $f_a$ for the QCD axion. The mass of the axion is $m$, and the potential flattens at $\phi\gtrsim M$ (see the left panel of Fig.~\ref{fig:Potential}). 

\subsection{Oscillon Solutions}
\label{sec:Osc_Sol}
In absence of couplings to other fields ($\g\rightarrow 0)$, the potential that we choose supports long-lived, spatially localized field configurations of the form 
\Beq
\phi_\omega(t,r)=\Phi_\omega(r)\cos\omega t+\hdots
    \Eeq
where $\hdots$ indicate, small higher order terms in harmonics of $\omega$ as well as an exponentially suppressed radiating tail. The solution is determined by specifying one parameter, $\omega$, which also determines the amplitude at the center $\Phi_\omega(0)$. For long-lived oscillon solutions, $\omega\lesssim m$ and $\Phi_\omega(0)\sim M$ (see the right panel of Fig.~\ref{fig:Potential}).

For the hyperbolic tangent potential, there is a special frequency $\omega_{\star}\approx 0.82m$, where the dominant decay channel for the oscillon vanishes, leading to exceptionally long-lived oscillons \cite{Zhang:2020bec}. Oscillons with lower frequencies (larger amplitudes) typically migrate towards this special configuration by radiating scalar waves, and tend to spend a significant fraction of their lifetime in this configuration. With these considerations, we chose oscillons with $\omega_{\star}\approx 0.82m$ for the pre-collision initial configurations of our axion fields before the mergers. For such oscillons
\Beq
\Phi_\omega(r=0)=2.4M\,,\qquad R_{1/e}\approx 3m^{-1}\,,\qquad E_{\rm osc}\approx 130 \frac{M^2}{m}\,,
\Eeq
where $R_{1/e}$ is the radius defined by $\Phi_\omega(R_{1/e})=e^{-1}\Phi_\omega(0)$, and $E_{\rm osc}$ is the total energy of the oscillon. Note that for a self-consistent treatment of oscillons with classical field theory, we want $E_{\rm osc}\gg m$, that is, $M^2/m^2\gg 1$.

Oscillons tend to be attractors in the space of solutions. The formation of such oscillons from cosmological initial conditions in the early universe has been explored in detail before \cite{Amin:2010xe,Amin:2010dc,Amin:2011hj,Gleiser:2011xj,Lozanov:2017hjm,Hong:2017ooe}. An almost homogeneous, oscillating condensate naturally fragments into oscillons. Late-time structure formation \cite{Levkov:2018kau}, or even nucleation near black-holes \cite{Hertzberg:2020hsz} could be a way of generating such configurations. The detailed investigation of formation, and merger rates is not undertaken in this paper. For this paper, we will take the existence of a pair of such objects as being given.

\subsection{Resonance Condition}
\label{subsec:RC}
\begin{figure}[t]
	\centering
	\includegraphics[width=1.01\linewidth]{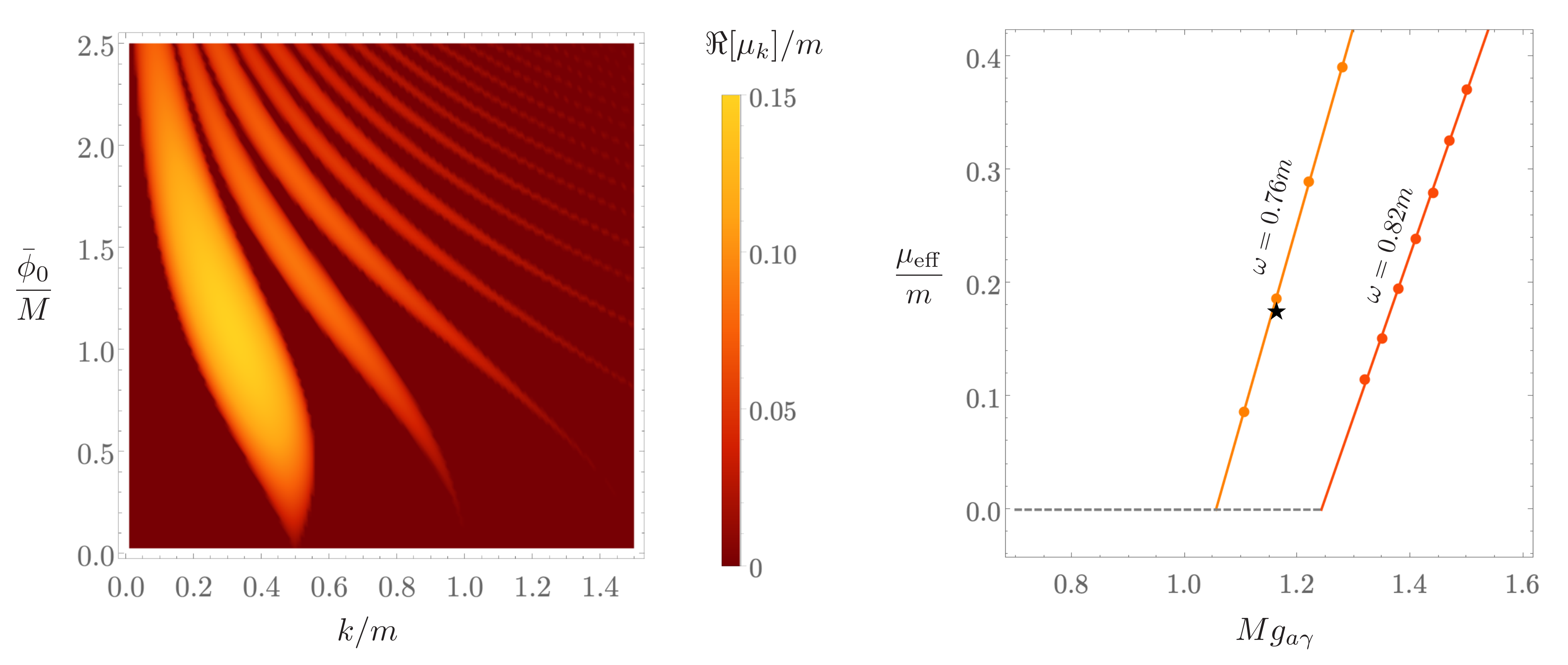}
	\caption{Left: The real part of the Floquet exponent of the gauge field in a homogeneous, oscillating axion field background. The vertical axis is the axion field amplitude and horizontal is gauge field wavenumber. The lighter regions are the unstable bands where the gauge field grows exponentially. We have assumed $M\g=1$ for the homogeneous plot. Right: The effective maximum Floquet exponent for gauge fields in spatially localized, oscillon backgrounds as a function of $M\g$. The luminosity in photons grows as $L_\gamma\propto e^{2\mu_{\rm eff} t}$. Note that $\omega=0.82 m$ and $0.76m$ correspond to two different oscillons (pre- and post merger configurations respectively). For each oscillon configuration, there is a critical value $(M\g)_{\rm crit}$ below which there is no parametric resonance into photons. For $M\g\approx 1.16$, the initial oscillons ($\omega=0.82m$) show no resonance, but post merger ($\omega\approx 0.76m$), there is resonant photon production with a Floquet exponent denoted by the star. Resonance is broad with $k_{\rm res}\sim \Delta k_{\rm res}\sim \mathcal{O}[1]\times\omega $, with some features seen at multiples of $\omega$.}
	\label{fig:Floquet}
\end{figure}

In this section we discuss the conditions necessary for resonant production of photons from individual oscillons which are spatially localized. As a warm-up, we first discuss resonance from a spatially homogeneous configuration. Parametric resonance from homogeneous oscillating condensates is well understood both in an expanding and non-expanding universe (see \cite{Amin:2014eta} for a review). 
A homogeneous oscillating background of the axion field (oscillating with a frequency $\omega\lesssim m$)  leads to a resonant transfer of energy from the axion to photons via parametric resonance. Explicitly, the equations of motion satisfied by the Fourier components of the gauge field (in Coulomb gauge) are
\Beq
\ddot{A}_{ik}(t)+\left[k^2+\g k\dot{\bar{\phi}}(t)\right]A_{ik}(t)=0\,,
\Eeq
where $\dot{\bar{\phi}}=-\omega\bar{\phi}_0\sin\omega t$ provides a periodic, time-dependent frequency. Then, from standard Floquet analysis, we expect the gauge field modes to have solutions of the form
\Beq
{A}_{ik}(t)\propto e^{\mu_k t}\,,
\Eeq
where the Floquet exponent, $\mu_k$, is a complex number which depends on the wavenumber $k$. For the case at hand, $\Re[\mu_k]\ne 0$, for a range of wavenumbers (see Fig.~\ref{fig:Floquet}, left panel). We have assumed $M\g=1$ for the purpose of illustration. Notice that this is quite different from the almost single band structure obtained in \cite{Hertzberg:2018zte}. This is related to our use of the hyperbolic tangent potential and $M\g\sim 1$.

Now if the axion field is spatially localized within a radius $R$ (like our oscillon), the resonance transfer of energy is still possible, as long as 
\Beq
\mu^{\rm eff}_k\approx \Re[\mu_k]-\frac{1}{2R}>0\,.
\Eeq
Heuristically, the size $R$ appears here because the growth of resonant modes is curtailed as they leave the region where the field amplitude is non-zero \cite{Hertzberg:2010yz,Hertzberg:2018zte}. 
For our initial oscillon with $\omega=\omega_\star\approx 0.82m$, the amplitude at its center is $\approx 2.4M$, and its width $R_{1/e}\approx 3m^{-1}$. Using these numbers, we find $\mu^{\rm eff}_k <0$. That is, there is no parametric resonance into photons for our fiducial oscillons according the above heuristic condition. We re-iterate that this is due to the finite size of the oscillon; a homogeneous condensate at this amplitude would transfer energy resonantly to photons.

A more careful analysis allows for spatial variation of the amplitude of $\phi$ within the oscillon, which in turn leads to coupling of Fourier modes. While this analysis can be done within some approximations \cite{Hertzberg:2018zte}, we do not pursue it here. Instead, in Fig.~\ref{fig:Floquet} (right), we show the numerically obtained $\mu_{\rm eff}$ as a function of $M\g$ for our oscillon with $\omega=\omega_\star$. This is obtained directly from the numerical simulation of an isolated $3+1$ dimensional oscillon coupled to photons. We take half the value of the exponent in the exponentially growing luminosity as $\mu_{\rm eff}$ ($L_\gamma\propto e^{2\mu_{\rm eff} t}$). We find that $\mu_{\rm eff}$ is (to a very good approximation), a linear function $M\g$ for $M\g>(M\g)_{\rm crit}\approx 1.24$. That is, we have exponential growth in gauge fields for any value of $M\g>1.24$, whereas below this value no exponential growth is seen. We note that while exponential growth in gauge fields is shut-off below $(M\g)_{\rm crit}$, there might still be polynomial growth which causes a very slow decay of the oscillon especially near $(M\g)_{\rm crit}$. Finally, we remind the reader that $(M\g)_{\rm crit}$ will depend on the oscillon profile. For comparison, in Fig.~\ref{fig:Floquet}(right), we also show $\mu_{\rm eff}$ for a different oscillon profile with $\omega\approx 0.76 m$. This oscillon has a larger amplitude and radius compared to the configuration with $\omega\approx 0.82m$.

\subsection{Time Scale of Resonance and Backreaction}
\label{sec:timescale}
If the oscillon configuration satisfies the resonance condition, then the energy in gauge fields increase with time $E_{\gamma}(t)\propto e^{2\mu_{\rm eff} t}$. The factor of two in the exponent is present because energy will be proportional to square of the gauge field. Then the time needed for the energy in gauge fields to become comparable to the oscillon is
\Beq
\label{eq:t_br}
mt_{\rm br}
&\sim \frac{m}{2\mu_{\rm eff}}\ln\left(\frac{E_{\rm osc}(t=0)}{E_{\rm \gamma}(t=0)}\right)
&\sim \frac{m}{2\mu_{\rm eff}}\ln\left(\frac{\rho_{\rm osc}(t=0)}{\rho_\gamma(t=0)}\right)
&\sim \frac{m}{\mu_{\rm eff}}\ln\left(\frac{M}{m}\right)\,.
\Eeq
In the second equality, we have limited ourselves to the core of the oscillon which is sensible since this is the region of production of the gauge fields. In the third line we used $\rho_{\rm osc}\sim m^2M^2$ and $\rho_\gamma\sim m^4$. Note that this would be the expectation from vacuum fluctuations, if we introduce a momentum cutoff at $\sim m$. Since these modes will become classical from resonance, it makes sense to include them in the initial energy density inside the confines of the oscillon. We can also use the energy density of photons from the CMB or starlight here depending on which one dominates at the mass and frequency of interest.\footnote{ Note that
\Beq
\rho^{\rm cmb}_\gamma(t=0)&=\frac{\omega^4}{\pi^2}\frac{1}{e^{\omega/T_{\rm cmb}}-1}\,,\\
\rho^{\rm sl}_\gamma(t=0)&\approx \frac{\omega^4}{\pi^2}\frac{1}{e^{\omega/T_{\rm
sl}}-1}W_{\rm dl}\,.\\
\Eeq
Note that $T_{\rm cmb}=2.3 \times 10^{-5}\,{\rm eV}$ today, and  $T_{\rm sl}\sim 0.43\,{\rm eV}$ for the dominant source for starlight in our galaxy today. The dilution factor for starlight, $W_{\rm dil}\sim 10^{-13}$ related to the distance between stars in our galaxy.} 
The above estimate assumes that the oscillon does not lose enough energy either via scalar fields, or gauge fields to fall out of the resonance condition before $t_{\rm br}$ and that $\mu_{\rm eff}$ remains constant. As we will see in later sections, oscillons fall out of resonance due to backreaction from gauge field emission before all of the oscillon energy is extracted by gauge fields. However, our estimate still serves as a useful guide for the time-scale involved.

\begin{figure}
\begin{center}
\begin{tabular}{lr}
\hspace{-2.2cm}
\includegraphics[width=0.7\textwidth,trim=0cm -0.8cm 0cm 0cm,clip]{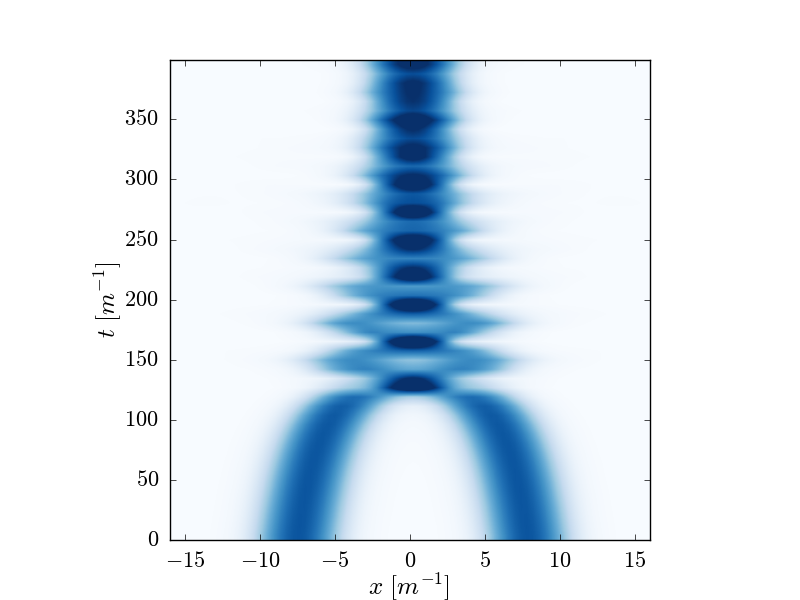} &
\hspace{-1.8cm}
\includegraphics[width=0.52\textwidth]{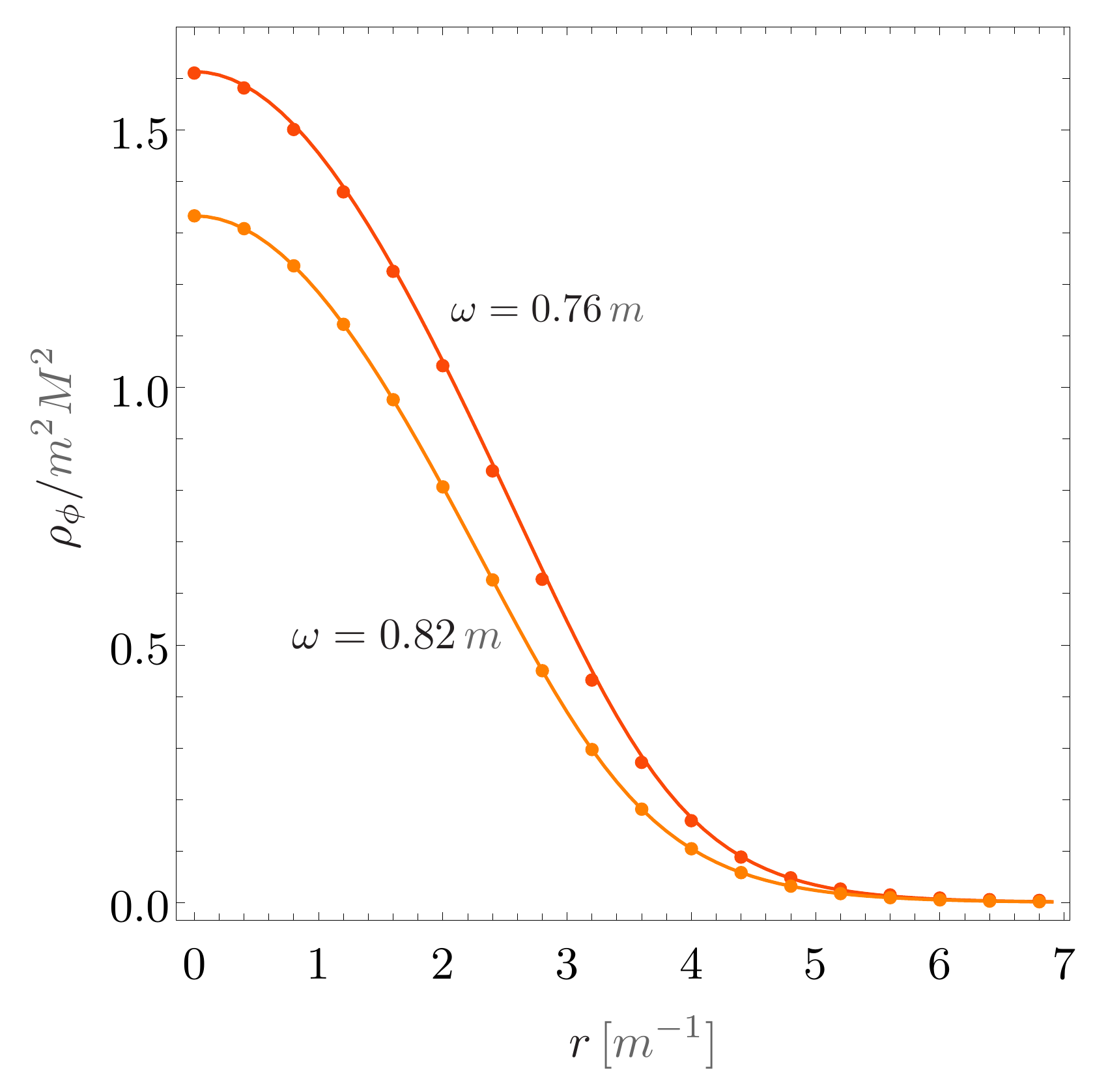}
\end{tabular}
\end{center}
\caption{
(Left) Energy density along the $x$-direction as a function of time (the simulations are 3+1 dimensional). The process of merger of two oscillons into a single oscillon, as well as emission of scalar radiation can be seen. Note that the merged oscillon is not quite in its ground state, it initially has a quadrupolar oscillating density pattern. The outgoing scalar waves are relativistic. The scalar radiation is almost $\sim 30\%$ of the initial total energy of the system, with the merged oscillon taking up the other $\sim 70\%$.  
(Right) The energy density profiles at different times for an oscillons, one before the merger, and one for post-merger. Solid lines are theoretical expectations, dots are extracted from our simulations (with time averaging).
}
\label{fig:en}
\end{figure}

\section{Collision and Merger of Oscillons}
\label{sec:Collisions}
We begin with two oscillons separated by some distance larger than their radii, moving towards each other with a small relative velocity $v\ll c$. For this section, we will ignore the coupling to gauge fields, and re-instate it in the next section. The two oscillons are assumed to be in phase, each oscillating with a frequency $\omega_{1,2}=\omega_\star$.

What is the end result of this collision? At least in the absence of gauge fields, we expect the oscillons to merge \cite{Amin:2019ums, Hertzberg:2020hsz} with each other, and form a new oscillon. A non-negligible fraction of the total energy of the oscillons is lost to scalar radiation. We find that
\Beq
E^{\rm (f)}_{\rm osc}\approx 0.7(E^{(1)}_{\rm osc}+E^{(2)}_{\rm osc})
\Eeq
That is, about $30\%$ of the initial energy of the field configuration (of two oscillons) is lost to scalar waves during the merger. For the post-merger oscillon, $E^{\rm (f)}_{\rm osc}>E^{(1,2)}_{\rm osc}$. For the model under consideration, this implies that the amplitude of the final oscillon will be larger than each initial oscillon, and its oscillation frequency $\omega_{\rm f}<\omega_{\star}$ (see left panel of Fig.~4 of \cite{Zhang:2020bec}).

This is indeed seen from our direct simulations of the collision and merger. For an initial separation between oscillon centers of $15m^{-1}$ and $v/c=10^{-2}$, we find that
\Beq
&\omega_{\rm f}\approx 0.76 m\qquad{\textrm{for}}\qquad \omega_{1,2}\approx 0.82m\,,\\
&E^{\rm (f)}_{\rm osc}\approx 180 \frac{M^2}{m},\qquad{\textrm{with}}\qquad E^{1,2}_{\rm osc}\approx 130\frac{M^2}{m}.
\Eeq
In Fig.~\ref{fig:en} (left panel), we show the energy density of axion field along the $x$ direction (which is the axis of collision), before and after the merger. The right panel shows the radial profile of oscillons before and after the merger, clearly indicating that the post-merger configuration is described well by an oscillon with $\omega \approx 0.76m$. Note that the amplitude and width of the merged oscillon is larger than the progenitors.

In reality this merger is dynamically richer. The oscillons can collide, separate a bit, and re-collide multiple times before finally settling down into a stable oscillon configuration. Moreover the merged oscillon is in an excited state initially, with a quadrupole density pattern which oscillates in time. The merger process includes emission of significant amounts of scalar radiation. We find that the amplitude of the merged oscillon decreases slowly. We also find that the oscillation frequency increases correspondingly so that the merged oscillon can be well described by adiabatically evolving oscillons after initial transients have subsided (and with some time averaging). Furthermore, the relative initial velocities will also impact the detailed merger process, with higher initial velocities leading to longer merger time-scales. For the case of head-on, in-phase collisions, relative velocities of up to $0.1c$ led to mergers.

We note that the fact the oscillons were in-phase, and identical is relevant for the end result being a relatively simple merger with negligible post-merger, center-of-mass velocity. If the initial oscillons are identical, but precisely out of phase by $\pi$, then they would ``bounce-off" each other. Small relative phase differences still lead to mergers, but lead to a small velocity for the final configuration \cite{Hertzberg:2020hsz}. We do not pursue different relative-phase, velocity possibilities in detail here, but focus on the in-phase merger case here because we expect it to be qualitatively similar to a broad swath of cases where the relative velocities are small, and the initial phase difference is not very close to $\pi$ \cite{Schwabe:2016rze,Amin:2019ums,Hertzberg:2020hsz}. If the initial relative velocities are ultra-relativistic, the oscillons would pass through each other \cite{Amin:2014fua}. For completely in-phase/completely out-of-phase collisions in  the strong field gravity regime, but without self-interactions, see \cite{Helfer:2018vtq,Widdicombe:2019woy}. 

\section{Resonant Electromagnetic Wave Production Post-Merger}
\label{sec:EMBurst}

In this section, we explore the production of electromagnetic radiation from the collision and merger of two oscillons. Recall that we have set up two oscillons moving towards each other with $v/c\ll1$. They are expected to be quiescent before merging, but produce a burst of electromagnetic radiation after merger. The merger and final nonlinear end stage of the coupled axion-photon system is hard to describe in detail from linear analysis, making detailed numerical simulations essential.

Before presenting and discussing the results of our simulations, we first provide some justification for our choice of two important parameters of the system.
\subsection{Choice of $M\g$ and $M/m$}
In the previous section we noted that the end state of the collision is another oscillon. We can then ask whether the resonance condition for gauge field production is now satisfied by the final oscillon. Recall that for the progenitors with $\omega_{1,2}=\omega_\star\approx 0.82 m$, we had $(M\g)_{\rm crit}\approx 1.24$. For the final oscillon, we found that $\omega_{\rm f}\approx0.76m$, and correspondingly  $(M\g)_{\rm crit}\approx 1.06$ (see Fig.~\ref{fig:Floquet}, right panel). Hence, for any $1.06<(M\g)<1.24$, we have a situation where there is no resonant production before merger, but there is resonant gauge field production post-merger. With this in mind, for concreteness, we take $(M\g)\approx 1.16$. We note that $\g\approx 1/M$ is larger than the expectation from QCD axions, where $\g\sim 10^{-2}M^{-1}$. However, recent theoretical work, such as \cite{Farina:2016tgd,Daido:2018dmu} allows for $\g M\sim 1$ in more general scenarios beyond the simplest QCD axion models, using for example the clockwork mechanism \cite{Choi:2014rja, Choi:2015fiu,Kaplan:2015fuy, Long:2018nsl, Agrawal:2018mkd}. For a survey of different mechanisms for getting a large $M\g$, see for example \cite{Dror:2020zru} and references therein.

Another parameter we need to choose is the ratio of $M/m$. As noted in Section \ref{sec:Osc_Sol}, $M/m\gg 1$ is necessary for us to trust our classical simulations of oscillons. As we will see in Section \ref{sec:Phenomenology}, $M/m\gg 1$ is also necessary from phenomenological considerations. For example, for typical QCD axions, we can have $M/m\sim 10^{25}$. However, from numerical considerations, such as the time scale required for the resonant production to extract significant fraction of the energy density from the oscillon (see eq.~\eqref{eq:t_br}), we cannot make $M/m$ so large. We will work with $M/m=10^4$ (but have also done simulations with $M/m=10^6$ and $M/m=10^8$). We argue below, that our results will be qualitatively similar to the cases when $M/m$ is much larger.
\begin{figure}
\includegraphics[width=\textwidth]{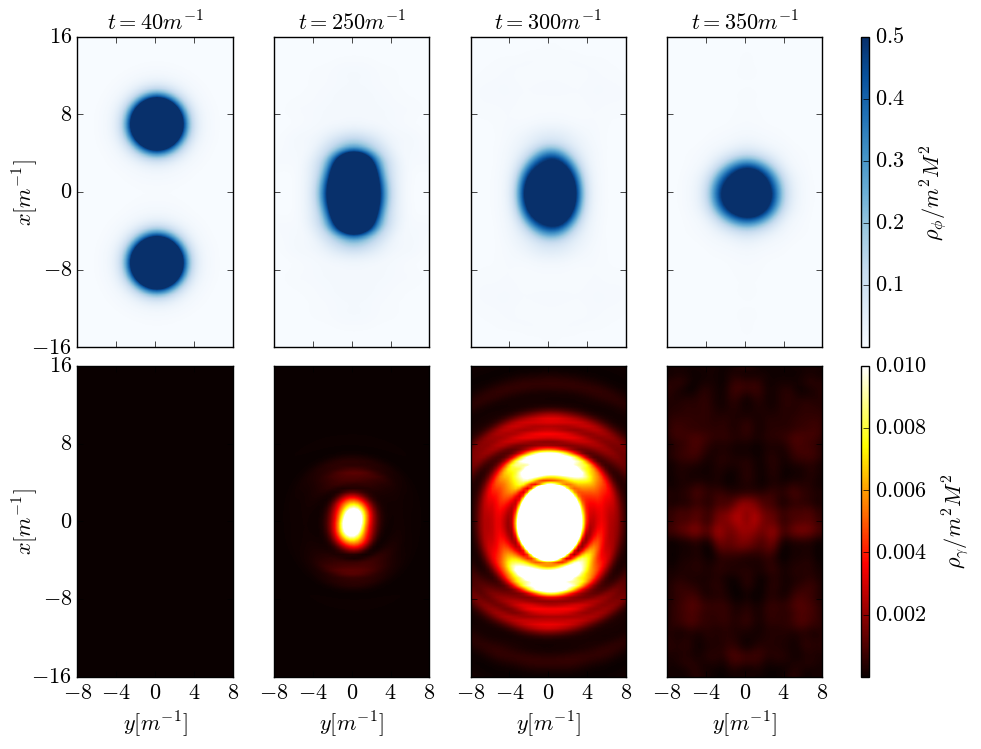}
\caption{The upper panels show the energy density of the axion field, whereas the lower panels show the energy density in the photons. Note that there is no photon production before merger. After merger, there is explosive (resonant) photon production which is eventually arrested again as the merged oscillon loses sufficient energy via photons to fall out of the resonance condition. Approximately $20\%$ of the merged oscillon energy is converted to gauge fields. Here,
$M/m=10^4$ and $Mg_{a\gamma}\approx 1.16$. We have checked that the above figure (including energy fractions) does not change qualitatively as we vary $M/m$ by a few orders of magnitude.  Although not visible in these snapshots, there is significant scalar radiation during the early stages of the merger ($\sim30\%$ of the initial total energy). Our simulation volume is more than double of what is shown in the snapshots with $L_x=77m^{-1}, L_y=L_z=51m^{-1}$.
}
\label{fig:decay}
\end{figure}
\subsection{Resonant Photon Production and its Backreaction}
The result of the oscillon collision and merger can be seen in Fig.~\ref{fig:decay}. The upper panel shows the energy density in the axion field, whereas the lower panel shows the energy density in the gauge field. As is evident from the panels, there is negligible photon production before the merger, but significant production after. In this simulation, the oscillons collide at $t\approx 120\,m^{-1}$.

\begin{figure}[t]
\begin{center}
\begin{tabular}{cc}
\hspace{-1.0cm}
\includegraphics[width=0.58\textwidth,trim=2cm 0cm 2cm 0cm,clip]{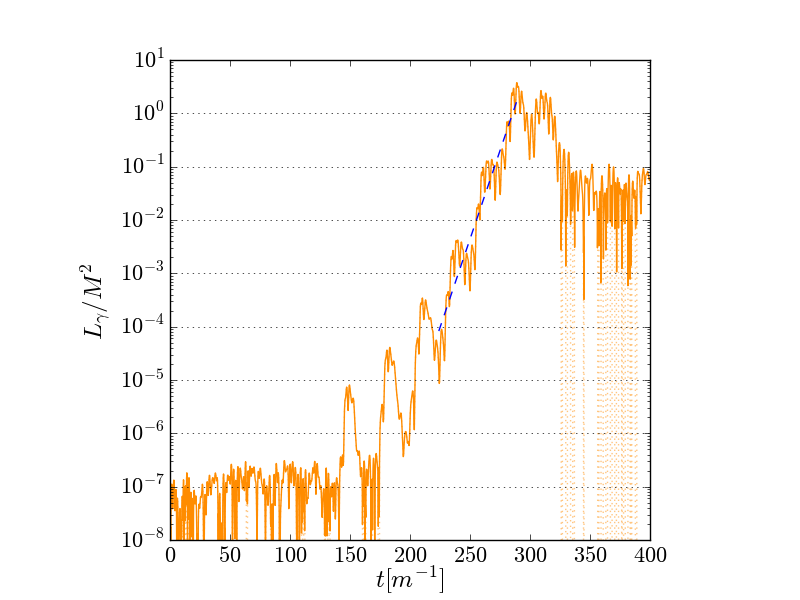} &
\hspace{-1.4cm}
\includegraphics[width=0.58\textwidth,trim=2cm 0cm 2cm 0cm,clip]{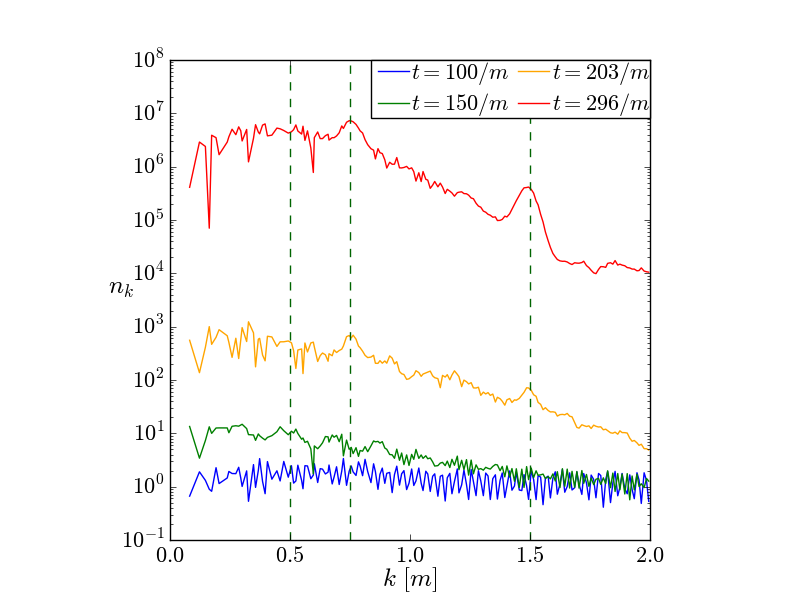}
\end{tabular}
\end{center}
\caption{
(Left) The exponential growth of photon luminosity after the merger for the case shown in Fig.~\ref{fig:decay}. The stopping of this exponential growth at $t\sim 300m^{-1}$ due to backreaction is also visible.  The effective Floquet exponent $\mu_{\rm eff}/m= 0.076$ can be inferred from the above plot. For this plot, $M/m=10^4$ and $Mg_{a\gamma}\approx 1.16$. The maximum value of luminosity in units of $M^2$ does not change significantly as we vary $M/m$ by two orders of magnitude or more -- the energy emitted is an approximately fixed fraction of the merged oscillon energy determined by backreaction considerations. The time-scale for backreaction only changes logarithmically with $M/m$, whereas the growth rate of luminosity is almost independent of $M/m$ for fixed $M\g$ as expected. See Fig.~\ref{fig:Lum_Comp} in the appendix for a comparison with the case where $M/m=10^6$, as well as other $M\g$. The luminosity at late times is affected by the radiation re-entering the simulation volume because of the periodic boundary conditions, in absence of which, the luminosity would plummet to small values. (Right) The time evolution of the occupation number in the gauge field. The resonance structure has a width of order $\Delta k\sim m$, with a peak around $k\approx \omega\sim m$ where $\omega$ is the frequency of the merged oscillon. The peak at $k=\omega\approx 0.76m$ and $k=2\omega$ is also visible (although the peak amplitude is not always monotonic in time). 
}
\label{fig:coll_lum}
\end{figure}

For a more quantitative picture, in Fig.~\ref{fig:coll_lum} (left panel), we show the luminosity (in photons). Notice that the luminosity only starts growing exponentially after the collision/merger. The exponential growth is well characterized by $L_\gamma \propto e^{2\mu_{\rm eff}t}$ with $\mu_{\rm eff}\approx 0.076m$. This value of $\mu_{\rm eff}$ is consistent with what is expected if the oscillon configuration after merger corresponds to $\omega\approx 0.76m$ (see Fig.~\ref{fig:en}). Note that after  $t\approx 300m^{-1}$, the exponential growth in luminosity stops. At this point oscillon configuration has radiated away $\sim 20\%$ of its initial energy density. Note that at this point, the oscillon configuration still exists, but has lost enough energy to gauge fields so that the resonance condition is no-longer satisfied. We expect the oscillon to eventually lose enough energy to return back to the $\omega\approx \omega_\star$ configuration and again spends a long time there, until another collision starts the process all over again.

The simulation results, in particular that the fraction of energy lost to photons is about $20\%$ of the energy of the merged oscillon, are shown for $M/m=10^4$. Since the shutting down of resonance is a backreaction effect with the oscillon configuration evolving away from the resonant domain, we expect this fraction to not change as we change $M/m$. We have checked explicitly, that this is indeed the case. We found that changing $M/m$ by two orders of magnitude (from $M/m=10^4$ to $M/m=10^6$) did not lead to any more than an order unity change in the energy fraction lost to gauge fields. We have also checked that the exponential growth rate of luminosity does not change significantly as we varied $M/m$, as expected. Furthermore, we have also verified that the time scale for backreaction is indeed logarithmic in the ratio $M/m$ (see Section~\ref{sec:timescale}).

The simulated behavior of gauge fields (and the system as a whole) at late times might be influenced by the finite size of our simulation volume which has periodic boundary conditions. As a result, one might worry that our simulation results might differ from infinite volume simulations where radiation truly leaves the system. In particular, in our simulation the luminosity does not quite drop to negligible values after the $t\sim 300m^{-1}$ because of the radiation coming back into the box which is unphysical. Ideally, we would like to significantly increased the simulation volume so than ${\rm min}[L_x,L_y,L_z]\gtrsim t_{\rm max}$, or implement absorbing boundary conditions. While our box size is smaller than $t_{\rm max}$, we have checked that changing the box size by a factor of $2$ did not effect the results qualitatively, at least up to $t\sim 300 m^{-1}$ or a bit longer. We have also made multiple checks by changing the resolution to make sure the growth rate of luminosity has converged.

While we do not show the results in detail here, we also simulated a case with $M/m=50$. In this case the backreaction completely destroys the oscillon. However, this destruction happens after several light crossing times for the simulation volume. As a result, we cannot be confident that the dynamics reflects the behaviour when the simulation volume is effectively infinite.

As mentioned in the previous section, there are additional transient dynamics during the collision (before the merger is complete). However these initial dynamics (transient overlap of the profiles) typically occur on a short time-scale compared to the eventual time-scale of gauge field production. The initial transient dynamics do lead to short bursts of gauge field production during the collisions, but the total energy released is subdominant compared to the final resonant release of energy. Nevertheless, observationally, such transient bursts might be interesting in confirming the origin of the electromagnetic burst signal.

\section{Phenomenological and Observational Considerations}
\label{sec:Phenomenology}

In this section we discuss some phenomenological and observational aspects relevant of our scenario. The discussion in this section is not as rigorous as the earlier sections, and details of some of the estimates re-derived here (for example, collision rates) can be found in more detail elsewhere.

If the axion makes up all of the dark matter, then
\Beq
\sqrt{m}M^2\sim m_{\rm pl}^{3/2}T_{\rm eq}\,.
\Eeq
Recall that $M$ plays the role of the familiar decay constant $f_a$, whereas $m$ is the mass of the axion-like field. We assumed that the energy density of the axion when it starts oscillating (ie. when $H\sim m$ in the radiation era)\footnote{For flattened potentials this is not necessarily true, and it is possible to have $H\ll m$ when oscillations begin (see, for example \cite{Kitajima:2018zco}). This will end up enhancing the hierarchy between $M$ and $m$ further.} is $m^2M^2$ and that it is equal to the radiation energy density at matter radiation equality. Using this constraint
\Beq
m\sim m_{\rm pl}^3 T_{\rm eq}^2M^{-4}=10^{-2}\,{\rm eV}\times\left(\frac{10^{12}\,\GeV}{M}\right)^{4}\,,
\Eeq
and the ratio $M/m$ is given by
\Beq
\frac{M}{m}\sim m_{\rm pl}^3T_{\rm eq}M^{-5}=10^{23}\left(\frac{M}{10^{12}\,{\rm GeV}}\right)^5\,.
\Eeq
The ratio $M/m$ large, unless $M\lesssim 10^8\,\GeV$. Note that we do not need to assume that the axion field is all of the dark matter, it could be a subdominant component. None of the results of the previous sections are affected. However, to reduce the available parameter space, we will take the axion to make up all of the dark matter.      
\subsection{Energy and Frequency of Emission}
The total energy locked up in our oscillons is 
\Beq
E_{\rm osc}\sim 10^2\frac{M^2}{m}\sim 10^{37}\,\GeV\times\left(\frac{M}{10^{12}\,\GeV}\right)^2\left(\frac{10^{-2}\,\rm eV}{m}\right)\,.
\Eeq
For the fiducial parameters, this is about $10^{10}$ kg, which is about the mass of the Pyramid of Giza locked in a radius $\sim 10^{-1}\,\rm cm$ (recall that $R\sim 5m^{-1}$).

The energy radiated into photons is about $\sim 10\%$ of this rest mass. Using the relationship between $m$ and $M$ from the dark matter abundance argument, we have
\Beq
E_\gamma\sim 0.1E_{\rm osc}\sim 10^{36}{\rm GeV}\times \left(\frac{M}{10^{12}{\rm GeV}}\right)^6\,.
\Eeq
This is a rather large amount of energy released in a short period of time from the collision of compact axion nuggets. It is comparable to the energy emitted by our sun in 1 sec. The electromagnetic radiation is emitted at a frequency and  bandwidth given by (see Fig.~\ref{fig:coll_lum})
\Beq
\omega_\gamma\sim \Delta\omega_\gamma\sim  m\sim 10^{-2}\,{\rm eV}\times\left(\frac{10^{12}\,\GeV}{M}\right)^{4}\,.
\Eeq
The time-scale associated with this energy emission can be estimated by the backreaction time (see Fig.~\ref{fig:coll_lum} or Section \ref{sec:timescale}):
\Beq
t_{\rm br}\sim 100m^{-1}\sim 10^{-11}\,{\rm s}\times\left(\frac{M}{10^{12}{\rm GeV}}\right)^4\,.
\Eeq
We have not included a logarithmic dependence on $M/m$ which can change this time by an order of magnitude.
Notice the strong dependence on $M$ of the frequency, energy and time-scale of emission; we will be rapidly shifting to lower frequencies, longer shorter time scales and larger energies as $M$ becomes larger.

Note that in the scenario envisioned here, we take $\g\sim M^{-1}$. The above expressions can be easily translated into those on $\g$. Current astrophysical constraints yield $\g\lesssim 10^{-10}\,{\rm GeV}^{-1}$ at $m\sim 10^{-2}\,\rm eV$, which translates to $M\gtrsim 10^{10}\,\GeV$. Detailed constraints from astrophysical sources and terrestrial experiments can be found in \cite{Tanabashi:2018ocaf}.

\subsection{Perturbative Decay}
So far we have not discussed the perturbative decay of axions to photons. This is simply because the perturbative decay time scale for $\phi\rightarrow \gamma+\gamma$:
\Beq
\Gamma_{a\gamma}^{-1}\sim 10^{2}(M\g)^{-2}\left(\frac{M}{m}\right)^2m^{-1}\,,
\Eeq
is typically large compared to the age of our universe. For $M\g\sim 1$, the large ratio $M^2/m^2$ controls the lifetime. For example, for $M\sim 10^{12}\rm GeV$ and $m\sim 10^{-2}\rm eV$, we have $\Gamma^{-1}_{a\gamma}\sim 10^{25}\rm yrs$, which only gets longer for lighter axions. 

\subsection{Binary Collision and Merger Rates}
We wish to estimate the number of collisions expected between dark matter clumps in a typical galaxy like ours. This collision rate can be estimated as follows (more details can be found in \cite{Hertzberg:2020dbk})
\Beq
\Gamma_{\rm coll} = \int dr 4\pi r^2\frac{1}{2}\left(\frac{f_{\rm osc}\rho_{\rm dm}(r)}{M_{\rm osc}}\right)^2\langle\sigma_{\rm eff} v\rangle\,,
\Eeq
where $\rho_{\rm dm}(r)$ is the smooth expected density of the dark matter halo, $f_{\rm osc}$ is the fraction of dark matter locked up in oscillons, $M_{\rm osc}$ is the mass of an oscillon, $v$ is the relative velocity between oscillons, the angled brackets imply a velocity average,  and $\sigma_{\rm eff}$ is the effective cross section for collision. This effective cross section is given by $\sigma_{\rm eff}=4\pi R_{\rm osc}^2\left(1+{v_{\rm esc,osc}^2}/{v^2}\right)$, 
where $v_{\rm esc,osc}^2/c^2=GM_{\rm osc}/R_{\rm osc}$. For a velocity average, $\langle\sigma_{\rm eff} v\rangle=\int_{0}^{v_{\rm esc}}  4\pi v^2 p(v)(\sigma_{\rm eff} v)$, we assume a distribution of the form $p(v)=p_0 e^{-v^2/v_0^2}$
where from normalization $p_0\approx (\pi/v_0^2)^{3/2}$ and $v_0=220\, {\rm km}\,{\rm s}^{-1}$ is the speed in the solar neighborhood. Note that the limiting velocity in the integral is the escape velocity for the dark matter halo, which we take to be $v_{\rm esc}=544\, {\rm km}\,{\rm s}^{-1}$. For simplicity, if we assume a constant density of dark matter up to a radius $R_{200}\equiv (3M_{\rm 200}/4\pi)^{1/3}$ with $M_{\rm 200}=10^{12}M_\odot$, then the binary collision rate within a galaxy like ours turns out to be 
\Beq
\Gamma_{\rm coll} \sim \mathcal{O}(1)\times \left(\frac{f_{\rm osc}}{10^{-2}}\right)^2\left(\frac{10^{12}\,{\rm GeV}}{M}\right)^4\left[1+ 10^{-6}\left(\frac{M}{10^{12}\,{\rm GeV}}\right)^2\right]\,\frac{{\rm collisions}}{{\rm galaxy}\,{\rm year}}\,.
\Eeq
This rate can be refined further, for example, by taking a more realistic $\rho_{\rm dm}(r)$ profile. Note that when oscillons from resonant instability in the axion field $f_{\rm osc}\sim \mathcal{O}[1]$. So, on the one hand we are being conservative here, by allowing for a much smaller fraction. However, since the lifetimes of oscillons might be shorter than the current age of the universe, this might be an overestimate. More generally, a detailed simulation of formation of halos (including oscillons) is desirable to get a more accurate estimate of the collision rate.

For head on collisions, we have found that mergers take place for relative velocities as high as $v/c\sim 0.1\gg v_{\rm esc}$, leading to expectations that even very high velocity collisions could lead to a merger. We suspect that the strong self-interaction has a significant impact on the probability of merger. Hence the merger rate might not be too different from the collision rate even when gravitational interactions are included, as long as the gravitational interactions are subdominant. The merger will of course be impacted by off-axis collisions, as well as relative phase differences between the solitons. There is likely an effect from the fluctuating ambient axion field in presence of which this merger takes place. For an argument regarding reduction in merger rates compared to collision rates in the context of dilute, gravitationally bound axion clumps (as opposed to our dense, self-interaction bound oscillons), see \cite{Hertzberg:2020dbk}. We leave a more detailed calculation of the true merger rate for our oscillons for future work. 

While non-electromagnetic signatures are not the focus here, constraints from gravitational signatures such as lensing from solitons (in the regime where they are sufficiently massive) can be further used to constrain the distribution of solitons, see for example \cite{Bai:2020jfm}.
\subsection{Observability}
The signal from a single event in our galaxy emits $E_\gamma\sim 0.1E_{\rm osc}\sim 10^{36}\,{\rm GeV}(M/10^{12}\,\GeV)^6$ amount of energy. This leads to a spectral flux density (flux/frequency bin): 
\Beq
S\sim \frac{E_\gamma}{\Delta \omega_\gamma \Delta t_\gamma(4\pi d^2)}\sim 10^8\left(\frac{M}{10^{12}\,\GeV}\right)^6\left(\frac{10\,{\rm kpc}}{d}\right)^2\,{\rm Jy}\,,
\Eeq
where $\Delta t_\gamma$ is the duration during which axions are rapidly converted to photons (and can be taken to be the backreaction time $t_{\rm br}$ estimated earlier), $\Delta\omega_\gamma\sim m$ is the band of frequencies of emission, and $d$ is the distance to the source. Note that $1\,{\rm Jy} = 10^{-26}\,{\rm Watts}/({\rm meter}^2 {\rm Hz})$. The steep dependence on $M$ comes from $E_{\rm osc}$ directly, with the $m$ dependence cancels in the product $\Delta \omega_\gamma \Delta t_\gamma$.  The spectral flux density is very high compared to the typical sensitivity of telescopes. Typically, this sensitivity will depend on the integrated time of the observations, but sensitivities below a Jy are not atypical. With this large $S$ for $d=10\,\rm kpc$, we can also hope to probe such events at cosmological distances $d\sim {\rm 100 Mpc}$. 

Recall that the frequency of emission is given by $\omega_\gamma \sim m\sim 10^{-2}\,{\rm eV}\left({10^{12}\,\GeV}/{M}\right)^4$. The fiducial frequency falls in the infrared regime, providing a potential target for James Webb Space Telescope (JWST), operating in a proposed survey mode \cite{Wang:2017awy}. Furthermore, the strong dependence on $M$ can be used to shift the frequency of emission. Depending on $M$, such events might be observable in any frequency range from gamma rays ($M\lesssim 10^{10}\,\GeV)$ to radio ($M\sim 10^{14}\,\GeV$). For $M\sim 10^{11.5}\rm eV$, we can get to optical frequencies, where they could become accessible to telescopes such as the Zwicky Transient Facility (ZTF) \cite{Dekany:2020tyb} or Vera Rubin Observatory \cite{Marshall:2017wph}. If the frequencies are in the radio regime, existing and future facilities such as the Canadian Hydrogen Intensity Mapping Experiment (CHIME)\cite{Andersen:2019yex}, the Long Wavelength Array (LWA) \cite{Lazio:2010dn}, the Low Frequency Array (LOFAR) \cite{vanHaarlem:2013dsa} and the Square Kilometer Array (SKA) \cite{Carilli:2004nx} might be able to detect such bursts. This very brief foray into observational aspects is rather shallow. A more careful assessment of detection possibilities is certainly worth pursuing. Moreover, a careful assessment of absorption and scattering of light (depending on the frequency) by the Intergalactic/Interstellar medium might also need to be taken into account \cite{Blas:2020nbs}.

It is intriguing that our spectral density can be made to match that of Fast Radio Bursts (for a review, see \cite{Petroff:2019tty,Platts:2018hiy}). A related mechanism, resonant radiation from collapsing axion miniclusters, was suggested as a source for Fast Radio Bursts in \cite{Tkachev:2014dpa}. Other related ideas in this context, for example axion-stars/oscillons falling on to neutron stars, include \cite{Iwazaki:2014wka,Prabhu:2020yif,Buckley:2020fmh}. 

\section{Limitations} 
\label{sec:Limitations}
There were a number of limitations to our study, which naturally point to future directions which can improve the present paper.
 
 We did not include gravitational interactions in our simulations. As a result, we could not explore low amplitude oscillons where gravitational interactions are needed to stabilize them (dilute axion stars). Note that distinct from earlier work, we worked in a regime where axion self-interactions dominate over gravitational interactions in potentials that flatten away from the minimum. Our oscillons, supported by attractive self-interactions, are long-lived in terms of their own oscillation timescales ($\gtrsim 10^7$ to even $\gtrsim 10^9$ of their own oscillations \cite{Zhang:2020bec}), however, they may not be long-lived compared to the astrophysical/cosmological timescale today.\footnote{The eventual decay of isolated oscillons is likely unavoidable, even in absence of coupling to other fields. Although the decay rates of oscillons is at times exponentially suppressed, the decay is still driven by the same attractive self-interactions which are necessary to hold the oscillon together in absence of gravity. Such considerations are of course always relevant when dealing with field theories with self-interactions, and in principle also present with gravitational interactions. In some cases the time scale of decay (especially deep in the perturbative regime), can be made comparable to astrophysical/cosmological time scales. Another possibility is that we can make the axions ultra-light, however, in this case the emitted electromagnetic radiation would not be detectable.}  We note that it is possible that including gravitational interactions might extend lifetimes significantly in some cases \cite{Fodor:2009kg,Eby:2015hyx, Eby:2017teq, Visinelli:2017ooc,Eby:2019ntd}. If lifetimes are short compared to cosmological time-scales, different formation mechanisms in the late universe (for example kinetic nucleation or nucleation near primordial blackholes \cite{Levkov:2018kau,Hertzberg:2020hsz,Kirkpatrick:2020fwd}) in addition to gravitational and self-interaction instabilities (for example, \cite{Amin:2019ums,Brax:2020oye,Arvanitaki:2019rax}) should be explored in detail to get a better understanding of the surviving population of solitons.  This is  something we have not addressed carefully in the present work and is worth pursuing in detail further. Our purpose here was to explore in detail the consequences of a collision between an existing pair of oscillons coupled to photons without worrying too much about how they got there.

Following some aspects of earlier work \cite{Eby:2017xaw,Hertzberg:2020dbk}, we calculated the collision rate using simple analytic arguments. We did not make a detailed effort to estimate the true merger rate.  Since we did not include gravity in our simulations, we also did not explore collisions from in-spiraling binary oscillons (or off-axis, out-of-phase collisions). Such effects could have an impact on the merger rate, and their respective time-scales which have to be characterized. A careful investigation in this direction is needed, and should include 3-body interactions and the presence of a spatio-temporally fluctuating axion background.\footnote{Some of the expectations are being pursued further in an ongoing collaboration on forces between oscillons, and oscillon collisions with Nabil Iqbal, Rohith Karur and Anamitra Paul.} We are making progress in this context in ongoing work.



\section{Summary}
\label{sec:Summary}
We investigated the production of scalar and electromagnetic radiation from the collision and merger of oscillons using 3+1 dimensional lattice simulations. We started with two identical, in-phase, oscillons, where they do not resonantly transfer energy to photons. As the collision and merger proceeds, initially there is a burst of scalar radiation. Then, if a resonance condition is satisfied, as the merged oscillon starts to settle, we get sustained resonant production of photons until the oscillon falls out of the resonance condition (due to energy lost to photons). To the best of our knowledge, this is the first time the entire process, including the backreaction, has been simulated. 

For our most detailed simulations, we chose $M\g\sim 1$ and $M/m=10^4$, where $m$ is the axion mass, $M$ is the scale where the potential becomes flatter than quadratic (similar to $f_a$ in the QCD axion case) and $\g$ controls the axion-photon coupling. We expect (and confirmed using $M/m=10^6$ and, in some aspects $M/m=10^8$) that our conclusions below will likely carry over for much larger $M/m$. We found that: 
\begin{enumerate}
\item During the collision and merger, about $\sim 30\%$ of the initial axion energy locked in oscillons is released as scalar radiation, where $E_{\rm osc}= \mathcal{O}[10^2]M^2/m$. 
\item About $\sim 20\%$ of the total energy of the merged oscillon is then resonantly transferred to photons, before the oscillon configuration changes sufficiently to shut off the resonance. The time scale for emission is $\sim 10\ln(M/m)m^{-1}$, and $E_\gamma \sim 0.1E_{\rm osc}$. The ratio $E_\gamma/E_{\rm osc}$ is approximately independent of $M/m$ as expected from resonance and backreaction considerations. This energy, $E_\gamma$, can be large enough to be detected by current and proposed telescopes over cosmological distances for some fiducial parameters.
\item The spectrum of energy of the emitted photons is centered around $\omega_\gamma\sim \omega_{\rm osc}\sim m$, however it is not exactly monocromatic. It has a width of order $\Delta\omega_\gamma\sim m$, with features related to multiples of the frequency of the oscillon resonantly producing the photons. The broad band structure, as well as the time evolution of individual modes is reminiscent of broad resonance.
\end{enumerate}
In detail, the merger and resonant photon production is dynamically complex, especially when backreaction is taken into account. For example, there are breathing and quadrupolar modes in the merged oscillon. Nevertheless, there are two crucial aspects for the phenomenology of resonant gauge field production from oscillon mergers: (i) The coherence of the field inside the merged oscillon is important for the resonant energy transfer to gauge fields. (ii) There is a threshold for resonant gauge field production in terms of the oscillon configuration for a give axion-photon coupling. As a result, parameter space exists where pre-merger oscillons do not efficiently transfer energy to gauge fields, whereas post-merger, as the threshold is crossed, we can get resonant photon production.

An important limitation of our work is that we ignored gravitational interactions. In absence of gravitational interactions, oscillons supported by attractive self-interactions alone will likely (but not necessarily) be short-lived compared to astrophysical and cosmological time-scales in today's universe (even if they last for billions of their own oscillations). Long life-times are possible in the dilute axion star regime, but in that case gravity plays a more prominent role compared to self-interactions. We believe that our oscillon merger and resonant gauge field production calculation is robust in the strong self-interaction regime. However, our result from the calculation for the rate of such events in a typical galaxy will be affected by formation, and survival probability of oscillons, which is much more uncertain. In upcoming work, we plan to include weak field gravity in the problem and carry out a more detailed calculation of the event rates. 

In this work, we only considered resonant production from axion-like fields to usual photons. However, we expect the phenomenology to work for dark photons also. As we discussed earlier a connection to Fast Radio Bursts would be worth exploring more carefully. Furthermore, our focus was on the contemporary universe. There might be related implications of photon/dark photon production from mergers of naturally abundant oscillons/solitons in the early universe.\footnote{However, if the couplings to gauge fields is sufficiently high, the initial branching ratio to oscillons from a homogeneous condensate might be suppressed.} Finally, we note that the emission of gravitational waves from collisions of oscillatons has been explored by some of us \cite{Helfer:2018vtq} and others in the past. It would be natural to combine that work with the present one (with appropriate parameter changes), to explore multi-messenger signals for such events.

\section{Acknowledgements}
MA is supported by a NASA ATP theory grant NASA-ATP Grant No. 80NSSC20K0518. We thank Mark Hertzberg, Andrea Isella, Mudit Jain, Andrew Long, Siyang Ling and HongYi Zhang for helpful conversations. 

\bibliographystyle{utphys}
\bibliography{reference}

\appendix

\section{Numerical Simulation Details}
\label{sec:AppNumerical}
The results discussed earlier are based on lattice simulations of the axion-gauge field system. Below, we provide details of our numerical algorithm as well as the initial conditions used.

\subsection{Equations of Motion}
The discrete equations of motion can be derived from the derivation of the action with respect to the fields, e.g. for the axion field
\begin{align}
\frac{ \phi(x+\ud t) - 2\phi(x) +\phi(x-\ud t)}{\ud t^2}=
\sum_i \frac{\phi(x+\ud x_i)-2\phi(x)+\phi(x-\ud x_i)}{\ud x_i^2}
-\frac{\ud  V}{\ud \phi}
+\frac{\delta S_1}{\delta \phi}
,
\end{align}
and for $U(1)$ gauge field
\begin{align}
\hspace{-0.7cm}
\frac{E_{i}(x)-E_{i}(x-\ud t)}{\ud t} =
- \sum_j\frac{i}{\ud x_i\ud^2x_j}\left[U_{ij}(x)-U_{ji}(x)+U_{ji}(x-\ud x_j)-U_{ij}(x-\ud x_j)\right]
-\frac{\delta S_1}{\delta A_i}.
\end{align}
In the equation above appears the lattice electric field, defined via 
$U_{0i}(x) - U_{i0}(x)  = -E_i(x)i\ud t\ud x_i$,
and accordingly the gauge link can be updated through
\begin{align}
U_i(x+\ud t) = \left[\sqrt{1-\left(\frac{1}{2}\ud t\ud x_iE_i(x)\right)^2} - i\frac{1}{2}\ud t\ud x_iE_i(x) \right] U_i(x).
\end{align}
To derive the expressions, we have adopted the temporal gauge, i.e. $A_0(x)=0$ or $U_0(x)=1$. 
Since $A_0(x)$ is not a dynamical field, the derivative of the action with respect to it leads to the constraint equation -- Gauss's law:
\begin{align}
\sum_i\frac{E_i(x)-E_i(x-\ud x_i)}{\ud x_i} =  \frac{\delta S_1}{\delta A_0},
\end{align}
which should be satisfied throughout the evolution.
Due to the smoothing scheme, there are many terms appearing in the interaction $S_1$, and we put these derivative of $S_1$ separately in Appendix \ref{sec:CS}.

\subsection{Initial Condition for Gauge Fields}
We choose the initial gauge fields that satisfy the following expectation values:
\begin{align}
\label{eq:expectation}
\langle A_i (\vec{p})A_j(\vec{k})\rangle &= \frac{n_p}{|p|}\left(\delta_{ij}-\frac{p_ip_j}{p^2}\right)(2\pi)^3\delta^3(\vec{p}-\vec{k}).
\end{align}
In practice, we assign a double of $n_p$ initially.
This is because we are constrained by Gauss's law, and not allowed to set both the field and their momentum expectation values arbitrarily. One way to satisfy the Gauss's law constraint is to set the momentum equal to zero. In that case, to have the same amount of energy, we choose to initialize the field by doubling the particle number.

On the lattice, the gauge fields are initialized explicitly via
\begin{align}
A_i(x) &=\sqrt{\frac{1}{V}} \sum_p e^{ipx} \sqrt{\frac{n_p}{2|p|}} \sum_{\lambda} \epsilon_i(p,\lambda)\xi(p,\lambda),
\end{align}
where random number $\xi(p,\lambda)$ satisfies $\langle\xi^*\xi\rangle=2$.
Only the physical photons are initialized, and their polarizations are set via
\begin{align}
\label{eq:polar}
\vec{\epsilon}(p,1)=\frac{\vec{r}\times \vec{p}}{|\vec{r}\times \vec{p}|}
,\quad
\vec{\epsilon}(p,2)=\frac{\vec{\epsilon}(p,1)\times \vec{p}}{|\vec{\epsilon}(p,1)\times \vec{p}|}
,
\end{align}
with $\vec{r}$ a random vector that is not parallel to $\vec{p}$.
By setting $A_i$ freely according to (\ref{eq:expectation}), we are no longer allowed to set the momentum fields $E_i$ freely, due to the Gauss's law.
In the non-Abelian case, one normally has to put $E_i=0$, while in Abelian case, like in our case now, we can instead set up $E_i$ by solving out the Gauss's law. 

\subsection{Initial Condition for Axion Field}

For the initialization of axion field, we consider two cases, depending on whether we are interested in exploring the resonance structure in a single oscillon, or simulating the collision and merger of two.

The single oscillon is initialized according to
\begin{align}
\phi(t,x,y,z) =& M f\left(\sqrt{x^2+y^2+z^2}\right) \cos\left(\omega t\right)
\end{align}
where $f(r)$ is the radial profile of oscillon.
We compute $f(r)$ in the following way.
For the Lagrangian, 
\begin{align}
{\mathcal L}=\frac{1}{2}\left(\partial_t\phi\right)^2 - \frac{1}{2}\left(\nabla \phi\right)^2 -\frac{m^2M^2}{2}\tanh^2\left(\frac{\phi}{M}\right)
,
\end{align}
we substitute the profile
$\phi = Mf \cos(\omega t)$
to obtain the action over one period
\begin{align}
\int_0^{2\pi/\omega} \ud t\int \ud^3x {\mathcal L}=\frac{ 4\pi^2M^2}{\omega}\int r^2\ud r \left[\frac{1}{2}\omega^2f^2 - \frac{1}{2}\left(\nabla f\right)^2 
-\frac{m^2}{2\pi} \int_{0}^{2\pi} \ud s \tanh^2\left(f\cos(s)\right)
\right]
,
\end{align}
from which, the equation of motion is straightforward,
\begin{align}
\partial_r^2  f + \frac{2}{r} \partial_r  f = -\omega^2  f 
+\frac{m^2}{\pi}  \int_{0}^{2\pi} \ud s \frac{\cos(s)\tanh\left(f\cos(s)\right) }{\cosh^2\left(f\cos(s)\right) }
.
\end{align}
Meanwhile, we can compute the average energy density over one period as,
\begin{align}
\rho_{\rm osc}(r)= \frac{M^2}{2}\left[\frac{1}{2}\omega^2f^2 + \frac{1}{2}\left(\partial_r f\right)^2 
+\frac{m^2}{2\pi} \int_{0}^{2\pi} \ud s \tanh^2\left(f\cos(s)\right)\right].
\end{align}

We also consider the collision of two oscillons, for which we initialize the axion field as 
\begin{align}
\phi(t,x,y,z) =& M f\left(\sqrt{\gamma^2(x-x_L-vt)^2+y^2+z^2}\right) \cos\left(\omega \gamma (t-v(x-x_L))\right)
\nonumber
\\
&+ M f\left(\sqrt{\gamma^2(x-x_R+vt)^2+y^2+z^2}\right) \cos\left(\omega \gamma (t+v(x-x_R))\right)
,
\end{align}
with the Lorentz boost $\gamma = 1/\sqrt{1-v^2}$ and $v$ the velocity of the oscillons.
The phenomenology of the collision between two oscillons is rich, with the phases, velocities and frequencies that one can vary.
For the present work, we limited ourselves to two oscillons of same phases and frequencies, but of opposite velocities.

\subsection{Numerically Evaluated Luminosity}
One of the observables that we relied upon heavily was the luminosity of photons produced by the resonant process. Luminosity in the continuum is of course defined as 
\begin{align}
L_\gamma=r^2\int \ud \Omega \frac{\vec{r}}{|r|}\cdot \left[\vec{E}\times \vec{B}\right]
.
\end{align}
Assuming the spherical symmetry, we can calculate the luminosity on the lattice, via 
\Beq
L_\gamma=\frac{4\pi r}{\mathcal{N}} \sum_{j=1}^{\mathcal{N}} \vec{r}_j\cdot [\vec{E}_j\times \vec{B}_j]\,,
\Eeq
where the sum is operated over sites of radius in $(r-\epsilon,~r+\epsilon)$.
In practice, we choose $mr=16$ and $m\epsilon=0.1$. After sufficient time, the luminosity is affected by the radiation that re-enters the central region because of periodic boundary conditions.
But, as long as the out-going radiation is much larger than the returning one, we can get the proper luminosity.
This is what happens to the resonance, for which we have tested out with different physical volumes and found the same $\mu_{\rm eff}$ during the exponential growth.

\begin{figure}[t]
\begin{center}
\begin{tabular}{cc}
\hspace{-1.0cm}
\includegraphics[width=0.6\textwidth]{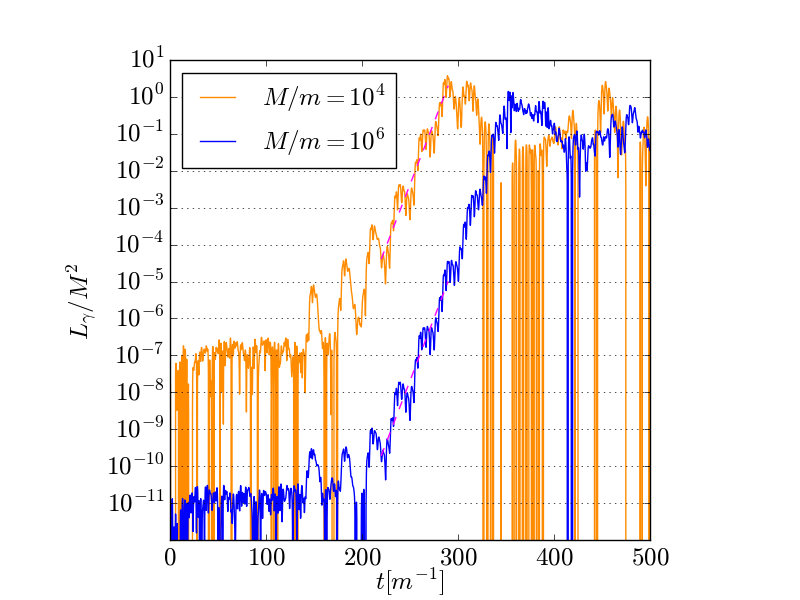} & 
\hspace{-1.4cm}
\includegraphics[width=0.6\textwidth]{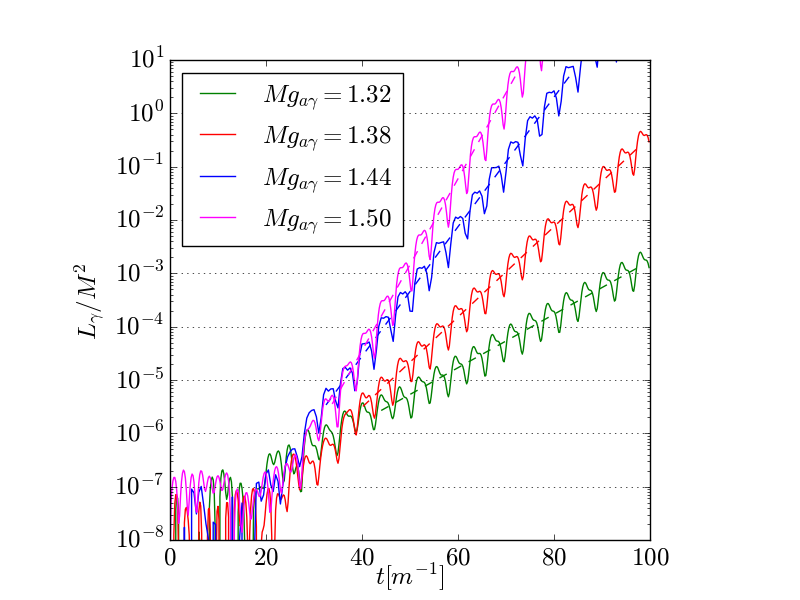}
\end{tabular}
\end{center}
\caption{
(Left)The exponential growth of photon luminosity after the merger for $Mg_{a\gamma}\approx 1.16$, and $M/m=10^4$ (orange) and $M/m=10^6$ (blue). The effective Floquet exponent $\mu_{\rm eff}/m= 0.076$ can be inferred from the plot for both cases, and is almost independent of $M/m$ as expected. The maximum value of luminosity in units of $M^2$ does not change significantly as we vary $M/m$. As seen above, the time-scale for backreaction only changes logarithmically with $M/m$. Note that the luminosity at late times (after exponential growth stops) is significantly affected by periodic boundary conditions. (Right) The exponential growth of the luminosity for different values of $M\g$.
}
\label{fig:Lum_Comp}
\end{figure}

\section{Chern-Simons Term on the Lattice}
\label{sec:CS}

For a general expression, we consider the Chern-Simons term,
\begin{align}
S_1= \int \ud^4 x \left[  -\kappa_1 \phi \left(-\frac{g_1^2}{64\pi^2}\right)\epsilon^{\mu\nu\rho\sigma}F_{\mu\nu}F_{\rho\sigma} \right]
,
\end{align}
with the $\phi-A$ coupling constant $\kappa_1$ and the gauge coupling constant $g_1$.
Then, the $U(1)$ Chern-Simon number admits
\begin{align}
\left(-\frac{g_1^2}{64\pi^2}\right)\epsilon^{\mu\nu\rho\sigma}F_{\mu\nu}F_{\rho\sigma}
= \left(-\frac{1}{2\pi^2 d^4x}\right)\left(I_{01}I_{23}+ I_{02}I_{31}+ I_{03}I_{12} \right)
,
\end{align}
with
\begin{gather}
I_{\mu \nu}(x) = \left(-\frac{i}{8}\right) \left[
U_\mu(x) U_\nu(x+\mu) U^\dagger_\mu(x+\nu) U^\dagger_\nu(x)
+U_\nu(x) U^\dagger_\mu(x-\mu+\nu) U^\dagger_\nu(x-\mu) U_\mu(x-\mu)
\right.\nonumber  \\
\left.+U^\dagger_\mu(x-\mu) U^\dagger_\nu(x-\mu-\nu) U_\mu(x-\mu-\nu) U_\nu(x-\nu)
+U^\dagger_\nu(x-\nu) U_\mu(x-\nu) U_\nu(x+\mu-\nu) U^\dagger_\mu(x)
-h.c
\right].
\nonumber
\end{gather}
To simplify the expression, we adopt the shorthand $(x+\mu) \equiv (x+\ud x_\mu)$ in the section.

The derivatives of the action with respect to different fields are straightforward, but tedious.
Here we include all these explicit expressions.

(i) Derivative with respect to $A_i$:
\begin{align}
\frac{\delta S_1}{\delta A_i}
=\frac{\kappa_1 g_1 dx_i}{16\pi^2 d^4x} \left(T_1[jk] - T_1[kj] - T_2\right)
,
\end{align}
with
\begin{align}
&T_1[jk]=
 \left[\frac{U_{ij}(x)+ U_{ji}(x)}{2} \right] \left[ \Xi_{0k}(x)+ \Xi_{0k}(x+i)+ \Xi_{0k}(x+j)+ \Xi_{0k}(x+i+j) \right]
\nonumber \\&
\hspace{-1cm}
-\left[\frac{U_{ij}(x-j)+ U_{ji}(x-j)}{2} \right] \left[ \Xi_{0k}(x-j)+ \Xi_{0k}(x+i-j)+ \Xi_{0k}(x)+ \Xi_{0k}(x+i) \right]
,
\end{align}
\begin{align}
\hspace{-5cm}
&T_2=
\left[\frac{U_{i0}(x)+ U_{0i}(x)}{2} \right] \left[ \Xi_{jk}(x)+ \Xi_{jk}(x+i)+ \Xi_{jk}(x+0)+ \Xi_{jk}(x+i+0) \right]
\nonumber \\&
\hspace{-1cm}
-\left[\frac{U_{i0}(x-0)+ U_{0i}(x-0)}{2} \right] \left[ \Xi_{jk}(x-0)+ \Xi_{jk}(x+i-0)+ \Xi_{jk}(x)+ \Xi_{jk}(x+i) \right]
,
\end{align}
where
\begin{align}
\Xi_{\mu\nu}(x) = \phi(x) I_{\mu\nu}(x)
.
\end{align}

(ii) Derivative with respect to $A_0$:
\begin{align}
\hspace{1cm}
\frac{\delta S_1}{\delta A_0}
=\frac{\kappa_1 g_1}{16\pi^2 d^3x}\left(T_3[ijk] + T_3[jki] + T_3[kij]\right)
,
\end{align}
with
\begin{align}
\hspace{-6cm}
&T_3[ijk]=
\left[\frac{U_{0i}(x)+ U_{i0}(x)}{2} \right] \left[ \Xi_{jk}(x)+ \Xi_{jk}(x+i)+ \Xi_{jk}(x+0)+ \Xi_{jk}(x+i+0) \right]
\nonumber \\&
-\left[\frac{U_{0i}(x-i)+ U_{i0}(x-i)}{2} \right] \left[ \Xi_{jk}(x-i)+ \Xi_{jk}(x)+ \Xi_{jk}(x-i+0)+ \Xi_{jk}(x+0) \right]
.
\end{align}

(iii) Derivative with respect to $\phi$:
\begin{align}
\frac{\delta  S_1}{\delta \phi}  = \frac{\kappa_1}{2\pi^2\ud^4x}
\left( I_{01}I_{23}+ I_{02}I_{31}+ I_{03}I_{12} \right) 
.
\end{align}

\end{document}